\documentclass{article}
\usepackage{graphicx}
\usepackage{amsmath}
\usepackage{enumitem}
\usepackage{tabularx,lipsum}
\renewcommand{\vec}[1]{\boldsymbol{#1}}

\usepackage[paperheight=11in,paperwidth=8.5in]{geometry}

\begin{document}
\title{A goodness of fit test for two component two parameter Weibull mixtures}
%\author{Author's Name}
\maketitle
\begin{center}
%\vspace{0.5 cm}

{\large Richard A. Lockhart}

{\small{\it{Department of Statistics and Actuarial Science, Simon Fraser 
University, Burnaby, B.C. V5A 1S6, Canada}}}

\vspace{0.5 cm}
{\large Chandanie W. Navaratna}

{\small{\it{Department of Mathematics,
The Open University of Sri Lanka,
Nawala, Nugegoda,
Sri Lanka}}}

\end{center}

\begin{abstract}
Fitting mixture distributions is needed in applications where data belongs to inhomogeneous populations comprising homogeneous subpopulations. The  mixing proportions of the sub populations are in general unknown and need to be estimated as well. A goodness of fit test based on the empirical distribution function is proposed for assessing the goodness of fit in  model fits comprising two components, each  distributed as two parameter Weibull.  The applicability of the proposed test procedure was empirically established using a  Monte Carlo simulation study. The proposed test procedure can be easily altered to handle  two component mixtures with different componet distributions.

\end{abstract}

\section{Introduction}

Fitting mixture distributions is needed in applications where data belongs to inhomogeneous populations comprising homogeneous subpopulations.  The  mixing proportions of the sub populations are in general unknown and need to be estimated as well. A goodness of fit test based on the empirical distribution function is proposed for assessing the goodness of fit in  mixtures comprising two components, each distributed as two parameter Weibull.  

Rest of the article is organized as follows. Section 2 describes mathematical formulation of the problem.  Section 3 illustrates the computation of the test statistic. Section 4 offers the asymptotic distribution of the proposed test statistic. Section 5 outlines a procedure for computing p-values based on the proposed test. Section 6 presents the results of a Monte Carlo simulation study that provides empirical evidence for the applicability of the proposed test procedure. Section 8 offers concluding remarks alone with a discussion.

\section{Two parameter Weibull mixture model and testing goodness of fit}

A random variable or vector $X$  is said to follow a finite mixture distribution, if the probability density function ( or probability mass function in the case of discrete $X$), $f(x)$ can be represented by a function of the form $f(x)=p_{1}f_{1}(x,\vec{\theta}_{1})+p_{2}f_{2}(x,\vec{\theta}_{2})+\cdots +p_{k}f_{k}(x,\vec{\theta}_{k}),$  where  $p_{i}\ge 0,$ for $i=1,2,\cdots,k$  are the mixing proportions such that  $\sum_{i=1}^{k} p_{i}=1$ and  $f_{i}(.) \ge 0$ are the density (or mass) functions of the components in the mixture such that  $\int_{\Omega} f_{i}(x) dx =1$  ( or in the discrete case $ \sum_{x \in \Omega} f(x)=1)$;  here  $\vec{\theta}_{i}$ denote the vector of parameters of the $i^{th}$ component density.
We assume that the mixture density is identifiable, so that for any two members $ \sum_{i} p_{i} f_{i}(x,\vec{\theta}_{i})=\sum_{j} p_{j}f_{j}(x,\vec{\theta}_{j}) ,$ if and only if  $p_{i}=p_{j}$ and  $f_{i}(x,\vec{\theta}_{i})=f_{j}(x,\vec{\theta}_{j}).$
In this work, we confine ourselves to identifiable mixtures with two components so that  $k=2$ and each component density is a two-parameter Weibull density given by 
  \[ f_{i}(x,\alpha_{i},\beta_{i}) =\frac{\alpha_{i}}{\beta_{i}} \left( \frac{x}{\beta_{i}}\right )^{\alpha_{i}-1} \exp \left(-\left( \frac{x}{\beta_{i}}\right)^{\alpha_{i}} \right) \] 

 The parameters $\alpha_{1}, \alpha_{2}$ are the shape parameters, $\beta_{1}, \beta_{2}$ are the scale parameters and  $\vec{\theta}_{i}=(\alpha_{i},\beta_{i})^{T}$ for $i=1,2.$  This model assumes that the location parameters of the two component densities to be the same.

 In this two component model, let  $p_{1}=p$ so that $p_{2}=1-p.$  Let  $F(x,\vec{\theta})$ denotes the mixture distribution function where  $\vec{\theta}=(\alpha_{1},\alpha_{2},\beta_{1},\beta_{2},p)^{T}$.
Given a random sample of n observations, from the distribution $F(x,\vec{\theta}),$    the goodness of fit problem  can be stated as a test of the null hypothesis that the distribution of the data is a two parameter Weibull mixture with parameter vector $\vec{\theta}$   that needs to be estimated in general. 

In the recent past, Weibull mixture models  have been extensively used in modeling wind data (\cite{akdag}, \cite{kollu}, \cite{sult}). In many of these studies, the goodness of the fitted models is examined based on Akaike Information Criteria (AIC), Basian Infromation Criteria (BIC), Chi squared test, Root Mean Squared Error (RMSE) and Kolmogorov Smirnov Test (K-S test). Sultan {\em et. al} \cite{sult} reports what they refer to as a correlation Goodness of Fit test for testing goodness of fit in mixtures of two Weibull distributions. In this work, we suggest a procedure for computing approximate p-values for testing goodness of fit of two component two parameter Weibull mixtures based on the Cramer-von Mises statistic. 

\section{Computation of the test statistic}

Let $F_{n}(x)$ denote the empirical distribution function  of the data defined by  $F_{n}(x)=\frac{1}{n} \sum_{i=1}^{n} I(x_{i} \le x), -\infty <x< \infty, $ where the indicator function $I(a,b)$  is defined as 1 for  $a \le b$ and as 0 otherwise.   Since $F_{n}(x)$  is the proportion of observations less than or equal to $x$  if $F(x)$  is the true distribution of  $X$ we expect $F_{n}(x)$ to be close to  $F(x).$
The closeness of $F_{n}(x)$  to $F(x)$  is assessed by the Cramer-von Mises statistics  defined by 
\[ W_{n}^{2}=n \int_{-\infty}^{\infty} (F_{n}(x)-F(x))^{2}dF(x). \] 
A computationally more feasible formula can be obtained by considering the probability integral transformation $z=F(x,\vec{\theta}).$ Let $x_{1},x_{2},\ldots,x_{n}$  be the order statistics of the original sample, then the probability integral transforms $z_{1},z_{2},\ldots z_{n}$  obtained as $z=F(x,\vec{\theta})$ will be an ordered sample of independent uniform[0,1] variables. If $\vec{\theta}$ is known, the test statistic can therefore be computed as  (see Stephens, Anderson[ ])
\[ W_{n}^{2}=\sum_{i=1}^{n} \left (z_{i}-\frac{2i-1}{2n}\right )^{2}+\frac{1}{12n}. \]

If $\vec{\theta}$ is not completely specified, and the null hypothesisis that the distribution is a member of the two parameter Weibull mixture distribution $F(x,\vec{\theta}),$ the same formula can be used to compute $W_{n}^{2},$  by using $z_{i}=F(x,\vec{\hat{\theta}}),$ where  $\vec{\hat{\theta}}$  is an asymptotically efficient estimate for $\vec{\theta}.$  In this work, we estimated $\vec{\theta}$ by the method of  maximum likelihood.

\section{ Limiting Distribution of the proposed statistic}
Literatuer reveals that (see Cramer[ ], Durbin[  ]) under suitable regularity conditions, the limiting distribution of  $W_{n}^{2}$ for testing the null hypothesis that $X$  is distributed as $F(x)$  is that of  $W^{2} = \sum_{j=1}^{n} \lambda_{j} z_{j}^{2}$ where  $z_{j}$ s are independent $N(0,1)$  variables and the  ’s are independent  variables and the  $\lambda_{j}$s are the eigenvalues of the covariance kernel $\rho$  namely, the solutions of the eigenvalue equation  $\int_{0}^{1} \rho(s,t)f(t)dt = \lambda f(s)$. It remains to discuss the computation of the eigenvalues of the covariance kernel. We present this separately for the two cases of simple hypotheses and composite hypotheses.

\subsection{ Simple Hypotheses}

 Durbin and Knott [ ] have proved that for simple null hypotheses, $\rho(s,t)$  is given by $\rho(s,t) = \min(s,t) - st.$  And, $\lambda$s can be computed in the closed form $\lambda_{j}=\frac{1}{\pi^{2}j^{2}}, j=1,2,\cdots .$ and the corresponding eigenfunctions are $ \sqrt{2}\sin(\pi_{j}s),$  for $j=1,2,\cdots,n.$

\subsection { Composite Hypothesis}
\label{eigQ}
In the case of a composite hypothesis,  $\rho(s,t)$  can be estimated by 
$\hat{\rho}(s,t) = \min(s,t) -st - \Psi(s)^{T} I^{-1} \Psi(t),$  where  $\Psi(s) = \frac{\partial F}{\partial \vec{\theta}} \left( F^{-1}(s,\hat{\vec{\theta}}),\vec{\theta}\right),$ where $\hat{\vec{\theta}}$   is an asymptotically efficient estimate for $\vec{\theta}$  and $I$ is the information of a single observation.

 Computation of the information matrix and inversion of the mixture distribution function are tedious for the Weibull mixture model at hand. We propose estimating the inverse of the information matrix using $-H/n,$ where $H$ is the Hessian matrix, or the matrix of second derivatives of the likelihood function evaluated at the maximum likelihood estimate $\hat{\vec{\theta}}.$ In passing we note that for the normal and exponential distributions, the matrix $-H/n$ gives the exactly correct form for the covariance kernel $\rho.$

\subsubsection*{Inverse of the mixture distribution function}
We propose computing the inverse of the mixture distribution function pointwise numerically. The procedure we used is described next. 

Given $t,$ we need to find $x$ such that $F(x,\vec{\theta})=t.$ This is equivalent to finding zeros of $g(x) = F(x, \vec{\theta})-t.$

We used Secant method, that gives the iteration scheme \[ x_{n+1} = \frac{x_{n}g(x_{n-1}) - x_{n-1} g(x_{n})}{g(x_{n})-g(x_{n-1})}; \]

here $g(x) = p \left( 1 - \exp \left( - \left(\frac{x}{\beta_{1}}\right)^{\alpha_{1}} \right) \right) + (1-p ) \left( 1 - \exp \left( - \left(\frac{x}{\beta_{2}}\right)^{\alpha_{2}} \right) \right)-t.$

The initial values needed to use this iterative scheme can be found by considering the boundary conditions for $p=0$ and $p=1.$

When $p=0,$ the condition $g(x)=0$ gives $ \left( 1 - \exp \left( - \left(\frac{x}{\beta_{1}}\right)^{\alpha_{1}} \right) \right) =t.$

Similarly, when $p=1,$ the condition $g(x)=0$ gives $ \left( 1 - \exp \left( - \left(\frac{x}{\beta_{2}}\right)^{\alpha_{2}} \right) \right) =t.$

Thus, $x_{1}=\beta_{1}   \log | (1-t) |^{1/\alpha_{1}}$ and $x_{2}=\beta_{2}   \log | (1-t) |^{1/\alpha_{2}}$  can be used as initial values. We note that since $t>0,$ $ \log(1-t)<0$ and hence it is essential to take the absolute value.

The iteration scheme can be carried out until desired convergence. We iterated until the difference between two consecutive points is less than a small number $\epsilon (>0)$ which we chosen to be $5 \times 10^{-6}.$ 
 
In all the examples we tried, the initial values for $x_{1}$ and $x_{2}$ obtained were on the opposite sides of the root and the iterative scheme worked satisfatorily.

To evaluate $\Psi(s),$ we need the derivatives $\frac{\partial F}{\partial \vec{\theta}}$ given by 
\begin{eqnarray*}
\frac{\partial F}{\partial \alpha_{i}} &= & p_{i} \left(\frac{x}{\beta_{i}}\right)^{\alpha_{i}} \log \left(\frac{x}{\beta_{i}}\right) \exp \left(-\left(\frac{x}{\beta_{i}}\right)^{\alpha_{i}} \right) \\ 
\frac{\partial F}{\partial \beta_{i}} &= &- p_{i} \left(\frac{\alpha_{i}}{\beta_{i}}\right)^{\alpha_{i}}  \left(\frac{x}{\beta_{i}}\right)^{\alpha_{i}} \exp \left(-\left(\frac{x}{\beta_{i}}\right)^{\alpha_{i}} \right)  \textrm{for} \quad  i=1,2  \quad  \textrm{and} \\
\frac{\partial F}{\partial p} &= &  \exp \left(-\left(\frac{x}{\beta_{2}}\right)^{\alpha_{2}} \right) - \exp \left(-\left(\frac{x}{\beta_{1}}\right)^{\alpha_{1}} \right). \\
\end{eqnarray*}

These derivatives have to be evaluated at $x=F^{-1}(s,\vec{\theta}),$ where $\hat{\vec{\theta}}$ is the maximum likelihood estimate.

Thus, at any point $(s,t)$ we can evaluate $\hat{\rho}(s,t). $ It remains to show how to calculate estimates for the eigenvalues of $\hat{\rho}(s,t). $ The eigenvalues of $\hat{\rho}(s,t)$ cannot be found in closed form and have to be estimated numerically.

\subsubsection*{Computation of estimates for the eigenvalues of the covariance kernel}

The difficulty associated with finding a closed form for the information matrix and inverting the mixture distribution function limits the application of methods proposed in the literature (see Stephens [5] and Stephens [6]) that hinges on the exapansion of $\Psi(s)^{T}I^{-1}\Psi(t)$ in a Fourier series in the eigenfunctions of $\rho(s,t)$ . We propose a brute force apprach for computing the eigenvalues that proceed as follows.

If $\lambda$ is an eigenvalue of $\rho(s,t)$ and $f(s)$ is an eigenfunction corresponding to $\lambda$, then $\lambda f(s) = \int_{0}^{1} \rho(s,t) f(t) dt. $

Divide the interval [0,1] into $(m+1)$ sub-intervals, each of which is of length $1/(m+1)$. Then,

\begin{eqnarray*}
\lambda f(i/(m+1)) & = &  \int_{0}^{1} \rho (i/(m+1),t) f(t) dt \\
& \approx & \frac{1}{m} \sum_{j=1}^{m} \rho(\frac{i}{m+1},\frac{j}{m+1})f(\frac{j}{m+1}), \quad \textrm{for sufficiently large $m$} \\
\end{eqnarray*}

Let $V$ be the column vector with $i$th element equal to $f(i/(m+1))$ and $Q$ be the $m \times m$ matrix whose $(i,j)$th element is $Q_{ij} = \frac{1}{m}\rho \left (\frac{i}{(m+1)},\frac{j}{(m+1)} \right).$

The above equation can be written as $  \lambda V = Q V.$ 
Hence, finding the eigenvalues of $\rho$ reduces to the dicretised problem of finding the eigenvalues of the matrix $Q.$

We developed software to create the matrix $Q$ using the estimate for $\rho(s,t)$ proposed in Section \ref{eigQ}.  Eigenvlaues of $Q$ were then used as estimates for $\lambda.$

\section{Computation of p-values}
\label{pval}

Having noted that the asymptotic distribution of $W_{n}^{2}$ is that of a weighted chi-squared distribution with eigenvalues of the covariance kernel as weights, approximate p-values can be computed as the probability $P(\sum_{i=1}^{\infty} \hat{\lambda}_{i} \chi_{1}^{2} \ge t), $ where $t$ is the value of the test statistic and $\hat{\lambda}_{i}$ are the estimates for the eigenvalues of $\hat{\rho}(s,t)$. We used Imhof's method \cite{imhof} to compute the approximate p-values.

Below we summarise the procedure for computing the suggested approximate p-values.

\begin{enumerate}
\item Find an asymptotically efficient estimate $ \hat{\vec{\theta}}$  of $\vec{\theta}=(\alpha_{1},\alpha_{2},\beta_{1},\beta_{2},p)^{T}.$  
\item Compute the probability integral transforms $z_{i}=F(x_{i},\hat{\vec{\theta}}). $
\item Compute the value of the test statistic, $W_{n}^{2} = \sum_{i=1}^{n} \left( z_{i}-\frac{(2i-1)}{2n}\right)^{2}+\frac{1}{12n}.$ \label{step3}
\item Compute $\hat{\Psi}(s)=\frac{\partial F}{\partial \vec{\theta}}(F^{-1}(s,\hat{\theta}),\theta),$ evaluated at $\hat{\vec{\theta}}$ at a desired grid of points $s$ in [0,1].
\item Estimate $I^{-1}$ by $- H/n,$ where $H$ is the Hessian matrix evaluated at $\hat{\theta}.$
\item Create the matrix $Q$ with $(s,t)$th element given by $\hat{\rho}(s,t)= \min(s,t)-st- \hat{\Psi}^{T}(s)(-H/n)\hat{\Psi}(t),$ for the grid points $s,t$ of the interval [0,1].
\item Find the eigenvalues $\hat{\lambda}$ of $Q.$
\item Compute approximate p-value as the probrability that the linear combination $\sum_{i=1}^{k}\hat{\lambda}_{i} \chi_{1}^{2} $ exceeds the test statistic value $W$ computed in Step \ref{step3}.
\end{enumerate}

\section{Empirical Justification for the proposed approximate p-value}
\label{simulation}

Weibull mixture populations for the simulation study were chosen to cover a range from poorly separated mixture components to well separated mixture components.  Table \ref{results} presents the parameters of the chosen mixture components. The results presented in Table \ref{results} is based on 10000 simulations for each. Approximate p-values for testing the composite hypothesis that the distribution is a member of the two component two parameter Weibull was computed using the procedure described in Section \ref{pval}.    If the procedure for computing approximate p-values is justifiable, the computed approximate p-values have to be uniformly distributed. The Anderson Darling test was used to examine the uniformity of the p-values. The last two columns of Table \ref{results} gives the values of the Anderson Darling statistic and the p-value for testing uniformity.

\begin{center}
\begin{table}[ht]
\begin{tabular}{cccccccc}
population & $\alpha_{1}$  & $\alpha_{2}$ & $\beta_{1}$  & $\beta_{2}$  & $ p$ &   Test statistic &   pvalue \\ \hline
  1 & 2 & 3 & 3 & 0.9 & 0.5 & 0.78 & 0.49 \\
   2 & 1.5 & 3 & 2 & 4 & 0.5 & 1.28 & 0.24 \\
3 & 1 & 3 & 2 & 4 & 0.5 & 0.95 & 0.38 \\
4 & 2 & 4 & 0.5 & 3 & 0.5 & 1.47 & 0.18 \\
5 & 2 & 8 & 1 & 4 & 0.5 & 2.43& 0.05  \\ \hline
\end{tabular}
\caption{Results of the simulation study}
\label{results}
\end{table}
\end{center}

None of the  results presented in Table \ref{results} provide evidence against the assumption that the resulting p-values are uniformly distributed. This can be taken as empirical evidence for the validity of the proposed test procedure fortesting goodness of fit in two component two parameter Weibull mixtures.

\section{Concluding Remarks and Discussion}

In this paper, we presented a goodness of fit test for  two parameter Weibull mixture models. Results of a Monte Carlo simulation study provided empirical evidence for the applicability of the suggested goodness of fit test. More simulation results are presented in Perera (\cite {perera}). Literature revealed applications of tests based on the Akaike Information Criteria and Bayesian Information Criteria (\cite {song}) as well as Root Mean Squared Error (RMSE), Chi Square tests, Kolmogorov-Smirnov test  (\cite {kollu}) in order to assess the goodness of fit in such mixture model fits. We expect the proposed test in this paper to be superior in terms of power; however, this needs to be established  using power studies against suitable alternative distributions. This is left as further work.

We also note that Likelihood surfaces of Weibull mixture distributions appear to be flat over a wide range in the parameter space. This gives rise to difficulties in calculating maximum likelihood estimates using simple procedures such as Newton Raphson method. Also, likelihood functions for samples of Weibull mixture densities that are not welll separated sometimes have more than one maximum; it is hard to find the global maximum with certainty. In such cases, we found that several very different roots can give equally good fits with similar likelihood values.

\end{document}